\documentclass[10pt,conference]{IEEEtran}

\usepackage{cite}
\ifCLASSINFOpdf
 \usepackage[pdftex]{graphicx}
\usepackage{amssymb}
\usepackage{lineno}
\usepackage{lscape}
\usepackage{tabularx}
\usepackage[gen]{eurosym}
\usepackage{multirow}
\usepackage{hhline}
\usepackage{booktabs}
\usepackage{siunitx}
\usepackage{float}
\usepackage{framed}
\else
\fi

\begin{document}
\title{What software engineering can learn from research on affect in social psychology}


\author{\IEEEauthorblockN{Lucas Gren}
\IEEEauthorblockA{Chalmers $|$ University of Gothenburg\\ Gothenburg, Sweden\\
Email: lucas.gren@cse.gu.se}
\and
\IEEEauthorblockN{Per Lenberg}
\IEEEauthorblockA{Chalmers $|$ University of Gothenburg\\ Gothenburg, Sweden\\
Email: perle@chalmers.se}
\and
\IEEEauthorblockN{Karolina Ljungberg}
\IEEEauthorblockA{SAAB AB\\ Gothenburg, Sweden\\
Email: karolina.ljungberg@saabgroup.com}
}

\maketitle
\begin{abstract}
Social psychology researchers have, traditionally, focused on the construct of thinking rather than on feeling. Since the beginning of the 21st century, social science researchers have, however, increasingly explored the effects of affect. Their work has repeatedly recognized that affects play a crucial role in determining people's behavior. In this short paper, we argue that software engineering studies on affect would benefit from using more of the knowledge that social science researchers have acquired. Without accounting for their findings, we risk re-inventing the wheel. Also, without a profound understanding of the complex interplay between social context and affect, we risk creating overly simplistic solutions that might have considerable long-term adverse effects for software engineers.
\end{abstract}
\IEEEpeerreviewmaketitle

%\section{Introduction}\label{sec:introduction}
\section{Why affect is essential to human life}\label{sec:life}
Affect plays a crucial role in the way people think and act in strategic social situations. It significantly influences the way people interact, formulate, and respond to interpersonal requests, plan and execute bargaining and negotiation strategies, and produce and use persuasive messages~\cite{forgas2002feeling}. A study on brain-damaged patients who lost their affective reactions, but retained their cognitive abilities, showed that such individuals make poor social decisions adversely influencing their social relationships~\cite{damasio1994descartes}. This indicates that humans are dependent on showing affect to navigate through social life, and social life is key to our survival.
% old: According to Forgas~\cite{forgas2002feeling}, affect plays a crutial role in the way people think and act in strategic social situations. Affect can: ``significantly influence the way people interact with each other, the way they formulate and respond to interpersonal request, the way they plan and execute bargaining and negotiation strategies, and the way we produce and use persuasive messages.'' A study on brain damaged patients who lost their affective reaction, but kept their cognitive abilities, showed that such individuals make disastrous social decisions that make their social relationships severely suffer \cite{damasio1994descartes}. This indicates that humans are dependent on showing affect in order to navigate through social life, and social life is key to our survival. %\cite{forgas2002feeling} gives further arguments for why affect is crucial for humans, but also relate it to some findings in biology in relation to other primates, which are described next.

Moreover, according to the prominent cognitive psychologist Pinker~\cite{pinker1997htm}, homo sapiens have been evolutionary successfully because of their ability to cooperate, interact, and influence others, which have helped us to survive and even prosper. Researchers have, also, shown that other primates are dependent on affect to navigate through social life using long-term emotional bookkeeping of reciprocal benefits~\cite{schino2009reciprocal}. Like other primates, humans have been shown to be affected by moods in expressing altruistic and helping behavior~\cite{altruhelp}. Humans social life is more complex than other primates due to their more advanced cognitive abilities and higher intelligence. In the modern world, humans life has changed. They have more interactions with strangers and superficially known others. Such interactions, which often are open and elaborate, have been shown to be more prone to affective influence~\cite{forgas2002feeling}. 
% old: According to the famous cognitive psychologist Stephen Pinker, homo sapiens have had the impressive evolutionary success because of our ability to cooperate, interact, and influence others, which have lead us to survive and even prosper \cite{pinker1997htm}. However, is has been shown that other primates are also utterly dependent on affect in order to navigate through social life through an emotional bookkeeping of reciprocal benefits that are not only immediate, but longer-term \cite{schino2009reciprocal}. Just like other primates, humans have been shown to be affected by mood in expressing altruistic and helping behavior \cite{altruhelp}. The complexity of social life is, of course, higher for homo sapiens due to much higher cognitive abilities and intelligence. Human life has also changed and in our modern world, we have more interactions with strangers and superficially known others, and such interactions that are open and elaborate have been shown to be more prone to affective influence \cite{forgas2002feeling}. 

\section{Connecting affect research in psychology to that of software engineering}\label{sec:connection}
Traditionally, social psychology research and, specifically, social cognition studies have focused on the construct of thinking rather than feeling~\cite{hogg2014sp}. An example of the latter is the study by Fiske, Cuddy, and Glick~\cite{fiske2007universal} in which they suggested that the two traits, warmth and competence, govern social judgments of individuals and groups, and that these judgments shape people's emotions and behaviors.

Since the beginning of the 21st century, researchers have increasingly started to focus on the effects of affect. The distinctions between different types of feelings (i.e.\ affect, emotions, and mood) are unclear though. Moods are commonly defined as more free-floating general feelings such as happy\slash sad, whereas emotions can often be attributed to some external cause~\cite{russell2003core}. What is clear, however, is that feelings influence, and are influenced by, social cognition~\cite{hogg2014sp}. Emotional responses to simple primary reactions of, for example, good\slash bad or harmless\slash dangerous are blindingly quick and occur in part of the brain called the amygdala~\cite{baxter2002amygdala}. The amygdala is part of the ``old brain,'' which developed early in our evolutionary history and is linked to autonomic system-related emotional reactions essential to our survival~\cite{russell2003core}.

% old: Traditionally, social psychology, and specifically, social cognition has focused on the construct of thinking rather than feeling \cite{hogg2014sp}. However, in the beginning of the 21st century many more researchers investigated the effects of affect. The distinctions between different types of feelings (i.e.\ affect, emotions, and mood) is far from clear, however, moods are more free-floating general feelings like happy\slash sad and emotions can often be attributed to some cause \cite{russell2003core}. What is clear, though, is that feelings influence, and is influenced by, social cognition \cite{hogg2014sp}. Emotional responses to simple primary reactions of, for example, good\slash bad or harmless\slash dangerous are blindingly quick and occur in the amygdala \cite{baxter2002amygdala}, which is the ``old brain'' linked to autonomic system-related emotional reactions that are essential to our survival \cite{russell2003core}. 

% The study conducted in software engineering has so far only included similar aspects.... %Skriva om SEmotion-artiklar... mapping study??? 

%  
% Similarly, Ortu et al.\cite{ortu2017connecting} affectiveness expressed by developers in their comments posted during the issue resolution phase in open source communities.
% Tools for extraction of empotions from text, for example open source comments~\cite{8595360}
% \cite{williams2017analyzing} tweets
% Jiarpakdee et al. found that using affective features improves the prediction of question quality~\cite{jiarpakdee2016understanding}. 
% quality

Moreover, Lerner and Dacher~\cite{lerner2000beyond} argue that valence, which is defined as the degree of attraction or aversion towards an object or event and often operationalized as good\slash bad mood, is a too simplistic division of the effect of emotions on judgment and choice. According to valence-based approaches, all negative emotions (e.g. fear or anger) will have the same effect on decisions. Still, Lerner and Dacher~\cite{lerner2000beyond} show that these two emotions have the opposite effect on risk assessments. Their work is based on the findings of Smith and Ellsworth~\cite{smith1985patterns}, who defined six cognitive dimensions that aggregate patterns of appraisal underlying different emotions: certainty, pleasantness, attentional activity, control, anticipated effort, and responsibility. Appraisal is defined as the act of estimating or judging the value or nature of something or someone, and the appraisal tendencies are ``goal-directed processes through which emotions exert effects on judgment and choice until the emotion-eliciting problem is resolved.'' By this definition, it is unclear what triggered the emotions, but the fact that they have a higher resolution than moods in their definitions implies that they are emotions and not moods. The distinction made between anger and sadness is that: ``anger arises from appraisals of individual control of negative events whereas sadness arises from appraisals of situational control of negative events.''~\cite{lerner2000beyond}. Lerner and Dacher~\cite{lerner2000beyond} conclude that negative emotions are likely to influence a variety of judgments in many different ways.

%old: However, in an article by \cite{lerner2000beyond}, they argue that valence (defined as the degree of attraction or aversion towards an object or event and often operationalized as good\slash bad mood) is a too simplistic division of the effect of emotions on judgment and choice. According to valence-based approaches, all negative emotions (like fear or anger) will have the same effect on decisions, however, \cite{lerner2000beyond} show that these two emotions have the opposite effect on risk assessments. They use the work by \cite{smith1985patterns} who defined six cognitive dimensions that aggregate patterns of appraisal underlying different emotions. These dimensions are: certainty, pleasantness, attentional activity, control, anticipated effort, and responsibility. Appraisal is defined as the estimating or judging the value or nature of something and someone, and the appraisal tendencies are ``goal-directed processes through which emotions exert effects on judgment and choice until the emotion-eliciting problem is resolved.'' By this definition, it is not clear what triggered the emotions, but the fact that they have a higher resolution in their definitions implies that they are emotions and not moods. The distinction made between anger and sadness is that: ``anger arises from appraisals of individual control of negative events whereas sadness arises from appraisals of situational control of negative events.'' \cite{lerner2000beyond}. \cite{lerner2000beyond} conclude that negative emotions are likely to influence a variety of judgments in many different ways. 

To summarize the distinction, moods are binary whereas emotions offer a higher resolution in their effects on behavior and are more subtle and long-lived which gives the possibility to investigate human life with more significant parts of affect influence~\cite{forgas2002feeling}. Emotions tend to disappear after the emotion-eliciting problem is resolved~\cite{lerner2000beyond}. People have, however, been shown to switch between moods, which makes that construct more volatile than was initially thought~\cite{forgas2002feeling}. 
%kl: Not clear. "which gives the possibility to investigate human life with more significant parts of affect influence"
% ---------------- Summary
The studies that have been published at the SEmotion workshop since its beginning in 2016 have been diverse in terms of research topics, as one could predict for such a young field of research. Roughly 30 papers have been accepted these years, most of which are based on the assumption that there is a positive link between positive emotions and productivity. For example, Graziotin et al.~\cite{graziotin2017consequences} shown that the unhappiness of developers negatively impacts productivity and performance. However, these results are somewhat contradicting studies in social psychology where a sad mood could sometimes be advantageous, e.g.\ \cite{forgas2007sad}, which then needs further investigation in the software engineering context to understand if and how it is different.

Moreover, many included papers have focused on the emotional content of comments and text in open source projects, such as GitHub, and tools for extracting emotion from such text have been analyzed~\cite{ding2018entity}. Related to the same topic, Marshall et al.~\cite{marshall2016outcomes} found that individuals with less emotive posts performed better than those that emoted more and that less affective individuals were evaluated more positively by their peers. Werder~\cite{werder2018evolution} showed that the team emotional display decrease over time in open source projects. Destefanis et al.~\cite{destefanis2018measuring} provided empirical evidence that commenters are less polite, less positive and in general, they express a lower level of emotions in their comments than users. We, however, note that several researchers have questioned the usefulness of such analysis tools since these have low agreement with humans' ratings~\cite{jongeling2017negative, murgia2018exploratory, 8595360}.

% End this paragraph with statement that ongoing studies focus 
Several studies have been related to how to create emotional awareness and identified monitoring mechanism for affect~\cite{condori2017using, liechti2017openaffect, sarkar2017characterizing, fountaine2017emotional, kuutila2018daily}. This information could, for example, be used to reduce the risk of ending up in an undesirable emotional state, such as stress~\cite{ostberg2017towards}.

Although we, by and large, appreciate such research efforts, we argue that the findings, if interpreted freely by practitioners, could have adverse effects on the psychosocial working environment. We see two potential risks that need to be addressed. Firstly, measuring and influencing emotions should not be considered an alternative to finding the root cause of problems. Clearly, the emotional state of human beings are highly complex, and researchers should be careful suggesting simplistic solutions to complex problems. Such attempts risk doing more harm than good. The study by Zhao et al.~\cite{zhao2017using}, in which they proposed to use playful drawing to elevate positive emotion, is an example of research that could be misinterpreted. Secondly, we question that it is desirable to strive for a working environment where positive emotions are the only acceptable state. Introducing an organizational culture in which negative emotions are considered unacceptable could put an unnecessary burden on employees forcing them to hide their true feelings. In such culture of silence, constructive debates and conflicts, needed to improve organizational processes, may be suppressed.

We argue that these two concerns, in general, should be more thoroughly discussed in the research papers and at workshops and conferences.

The division between emotions and moods in social psychological research implies a corresponding set of research questions. The emotion-specific influences are useful when investigating the effects of all basic emotions or core affects (see~\cite{russell2003core}), which would then have to be pinpointed to the causes of the emotions. Mood research, on the contrary, offers an awareness of more subtle effects to which more general interventions of increasing happiness can be introduced. 

The research on emotions in software engineering has thus far focused on individual emotions without accounting for context, and has, therefore, committed the same initial journey as social psychology. However, we must not re-invent the wheel, but can instead learn from in what ways social psychology has proceeded in its research. It is not sufficient to state that emotions influence behaviors and look for general actions to change the mood; instead, we argue that the root causes need to be treated instead of the symptoms. % Software engineering researchers have, thus far, often focused on developer happiness~\cite{graziotin2014happy}.

Before context was accounted for in social psychology research, two preceding affect theories were the affect-as-information approach and the affect priming theory. For situations where no elaborate processing strategy is needed, one heuristic known to be used is to try to find information in a feeling towards an object. People mistakenly think that a feeling is a reaction to the target, which might as well not be the case, i.e.\ for a quick assessment people ask themselves how they feel in relation to the object. 
%pl: I don't understand what you mean here (the last sentence). Glasklart ju.
%kl: Might be obvious if you already know the theories but I didn't understand that the the affect-as-information approach was discussed until the next paragraph started with "In contrast, the affect priming theory..."
In contrast, the affect priming theory explains situations of more elaborate processing by assuming that affect is ``an integral aspect of cognitive representations about the social world''~\cite{forgas2002feeling}. Since these theories have produced somewhat inconsistent results, Forgas~\cite{forgas2002feeling} instead presumes that the social influences depend on which information-processing strategies people choose in context. We will now explain how context is accounted for in the Affect Infusion Model (AIM). 

\begin{figure}
\centerline{\includegraphics[width=90mm]{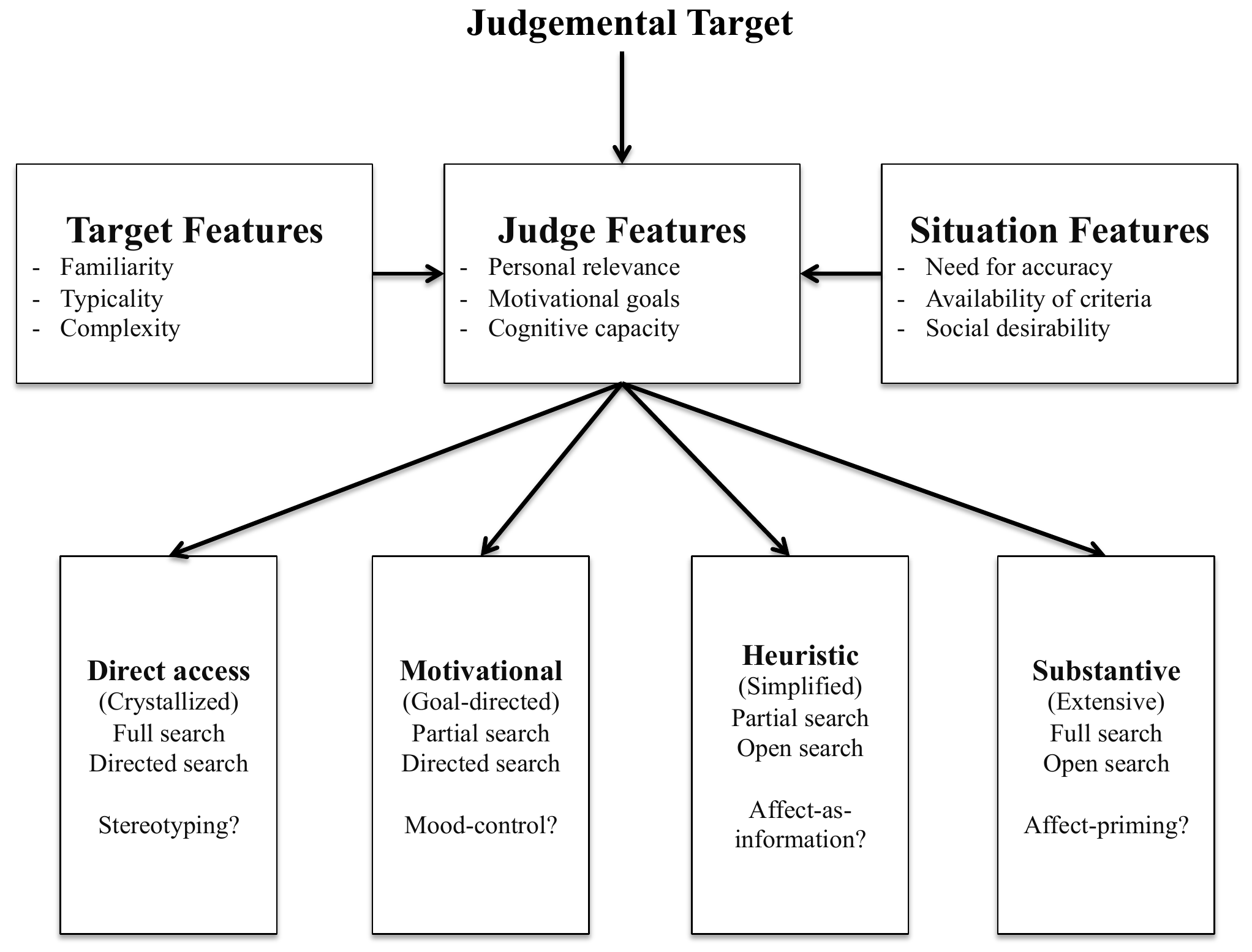}}
\caption{The AIM predicts that direct access and motivated processing are little influenced by mood, whereas heuristic and substantive processing are highly influenced by mood in judgments \cite{forgas1995mood}.}
\label{aim}
\end{figure}

Forgas~\cite{forgas2002feeling} argues that context had been largely ignored before his article in 2002 and that behavioral psychologists have, therefore, contributed very little to the understanding of affective influence. He suggests the Affect Infusion Model (AIM) that attempts to explain why affect-congruence, affect-incongruence, or no effect at all are found in relation to affective influences of interpersonal behavior. Through the AIM, Forgas~\cite{forgas2002feeling} explains the previous findings in two distinctly different dimensions, i.e.\ the processing strategy is determined by the degree of effort needed and the degree of openness and constructiveness. With two alternatives on each dimension, we obtain the four different processing strategies: direct access processing (low effort, closed and not constructive), motivated processing (high effort, closed and not constructive), heuristic processing (low effort, open and constructive), and substantive processing (high effort, open and constructive). See Figure~\ref{aim} for details about the four different strategies for social judgment \cite{forgas1995mood}. The AIM describes the features of the target, the features of the judge (the person making the decision), and the situational features of the surrounding environment. The AIM then describes the four different strategies that can be selected.

According to the model, affect is the most likely an influence when an open and constructive strategy is chosen, which means that a substantive processing style should lead to high affect infusion. This is an interesting proposal that also has gained empirical evidence~\cite{forgas2002feeling} and moves focus from the previously assumed effort dimension to the openness and constructiveness dimension when understanding affect infusion. Therefore, more complex, demanding, and novel tasks, which are of high effort but with an open-ended information search, are then prone to affect infusion~\cite{forgas1995mood}. The reason why Figure~\ref{aim} includes question marks is that the model was developed before the different predictions had been tested empirically.  

We argue that modern software development teams are in such contexts, i.e.\ most people involved in decision-making in the software engineering context need to apply substantive or heuristic reasoning because they often need to be creative. Therefore, research on affect in software engineering could make use of the AIM in future research in order to navigate through different affective responses, but, and perhaps more importantly, to add the contextual features of the target, the judge, and the situation. 

In addition, affect itself can influence which processing strategies that are chosen, which is not depicted in Figure~\ref{aim}. Positive affect has been shown to generate a more top-down and heuristic processing~\cite{hertel1994affective} while a sad mood has been shown to trigger a more bottom-up and vigilant processing~\cite{forgas2007sad}. A sad mood has even been shown to reduce the fundamental attribution error (i.e.\ the tendency to attribute another person's behavior more to internal than to situational causes)~\cite{forgas1998being}.

An interesting and somewhat counter-intuitive finding is that happy people prefer more direct and impolite requests~\cite{rudehappy}. This result confirms the tendency to use a more heuristic processing style when in a happy mood. Because of these same mechanisms, some studies have also shown that a happy mood can lead to the selection of more cooperative strategies in negotiation~\cite{forgas1998feeling}, which is, of course, often an advantage. Here we see great potential for research in requirements engineering. In the more general organizational context, Sinclair~\cite{sinclair1988mood} showed that depressed people form more accurate appraisals that are less subjected to primacy effects, which are order-of-presentation effects where earlier presented information has a disproportionate influence on social cognition (sometimes also known as halo\slash horns effects).

%old: An interesting and somewhat counter-intuitive finding is that people in a happy mood prefer more direct and impolite requests \cite{rudehappy}, however, this result confirms the tendency to use a more heuristic processing style when in a happy mood. Because of these same mechanisms, some studies have also shown that a happy mood can lead to the selection of more cooperative strategies in negotiation \cite{forgas1998feeling}, which is of course, many times an advantage. Here we see great potential for research in requirements engineering. In the more general organizational context, \cite{sinclair1988mood} showed that people in depressed moods form more accurate appraisals that are less subjected to primacy effects.\footnote{An order of presentation effect where earlier presented information has a disproportionate influence on social cognition. Sometimes known as halo\slash horns effects.}

None of these dimensions and findings have, however, been used or replicated in software engineering. We argue that using models that include context is a necessity for reaching practical significance in research on affect in software engineering.
%None of these dimensions and findings have been used or replicated in the software engineering case, and we argue that using such model that include context, is a must to reach practical significance in research on affect in SE.

To summarize the research on affect in social psychology, a positive mood may help humans overcome defensiveness and better deal with threatening situations~\cite{forgas1998feeling}. In addition, people shift between moods and people in a negative mood become spontaneously more positive over time, which implies that moods are highly dynamic~\cite{sedikides1994incongruent}. 
Again, none of these dimensions or perspectives have been included in any research on affect in software engineering.

\section{Conclusion}\label{sec:conclusion}
Drawing on findings from social psychology research on emotions, we argue that controlling software engineers' moods might have considerable long-term adverse effects for employees. Social dynamics are seldom accounted for in software engineering affect research, which introduced a high risk of developing interventions of the symptoms of being unhappy instead of the root cause. Previous research show that the social context is essential for understanding affect~\cite{forgas2002feeling} and that humans are affected by mood in expressing altruistic and helping behavior~\cite{damasio1994descartes}.
%old: Relating the results from social psychology to affect research on software engineers, we believe controlling moods is futile and might, in the worst case, lead to pathetic, happiness-inducing, and circus-like events in organizations. Context has been shown to be essential to understanding affect~\cite{forgas2002feeling}, and humans have been shown to be affected by mood in expressing altruistic and helping behavior~\cite{damasio1994descartes}. Social dynamics has so far been left out in research on affect in software engineering, which introduced a high risk of developing interventions of the symptoms of being unhappy instead of the root cause.

Social psychologists have developed questionnaires that include context and social relationships, which can directly be used in software engineering research. One example of such a questionnaire is the mapping of social identity through a collective self-esteem evaluation (using the scale proposed by Luhtanen and Crocker~\cite{luhtanen1992collective}). If we want to increase the productivity in software development organization, we might succeed better if we conduct interventions to increase the collective self-esteem instead of trying to directly change their mood.
%pl: I tried to understand what you meant, but my sentences does not make any sense.
%old: There are questionnaires that have been developed in social psychology that include context and social relationships, which can directly be applied to software engineering research. One example of an interesting questionnaire to use could be the mapping of social identity through a self-esteem evaluation (using the scale proposed by~\cite{luhtanen1992collective}) and connect an intervention to increase self-esteem, might be more useful if the end-goal is to have employees with more stable and high productivity despite their different moods. 

\bibliographystyle{IEEEtran}

\bibliography{references}

\end{document}